# Abstract

*Dejan Grba*

Incidental Reverberations: Poetic Similarities in AI Art

**Keywords** | *AI art, Creative thinking, Poetic similarities, Computational art*

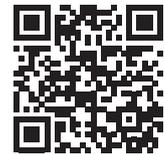

*Contemporary AI art's diverse and widely recognized repertoire features numerous artworks that share conceptual, thematic, narrative, procedural, or presentational properties with other artworks across disciplinary and historical spectrums. AI artists occasionally leverage well-sanctioned poetic referencing as an asset but when obvious or easily discoverable similarities remain unacknowledged, they may become liabilities. Lurking behind the hype waves in the media, art world, and academia, these liabilities shape contemporary AI art's cultural identity and affect its social impact. As part of a broader study of poetic contingencies, in this paper I discuss selected AI art exemplars whose multifaceted expressive parallels are symptomatic of the field and beyond. I argue that expressive similarities in AI art are as detrimental to its cultural value as they are avoidable in its variety of important topics addressable with a wide range of creative affordances. My critique takes the well-informed autonomy of expression and the socially responsible freedom of creative thinking as the tenets of artmaking to indicate some of AI art's related issues and challenges induced by its entanglements with AI science, technology, and industry. In conclusion, I suggest that poetic similarities open a valuable perspective for studying AI art's strengths and deficiencies and for articulating a broader critical discussion of art and creativity.*

*Dejan Grba*  Incidental Reverberations: Poetic Similarities in AI Art

## 1. Introduction

Since its outset in the 1970s, AI art remained largely obscure until the expansion of digital infrastructures and the availability of versatile computational tools in the early 2000s facilitated its poetic diversification. The confluence of increasingly accessible machine learning (ML) techniques and the rising social impact of AI since the 2010s, have allowed it to gain productive momentum, cultural presence, the AI industry's patronage, and the art market's recognition.[1] Contemporary AI art features experimental, exploratory, and speculative practices that emerge from and respond to modern AI's phenomenological realms and social implications.[2] However, these diverse practices often result in artworks that share conceptual, thematic, narrative, procedural, or presentational properties with other artworks across disciplinary and historical spectrums. While some AI artworks leverage well-sanctioned poetic referencing as an asset, in cases where obvious or easily discoverable similarities remain unacknowledged, they may become liabilities.

Lurking behind the hype waves in the media, art world, and academia, these "incidental reverberations" frequently happen because various traits and drives, such as audaciousness and calculated ambitions, compel or lure artists to disregard the extent and open-endedness of informed external interpretation and intervention. They belong to a corpus of intellectual blunders and methodological miscalculations whose unforeseen consequences are usually undesired by the artists but always instructive for their audience. They affect AI art's cultural identity and social impact by raising concerns about the artists' contextual awareness of current art production, art-historical knowledge, or professional ethics. As part of a broader critical study of poetic contingencies in computational art,[3] this paper summarizes the issue of widespread expressive similarities in AI art by discussing selected exemplars whose multifaceted parallels are emblematic of the field.

## 2. Incidental Reverberations

Conceptual parallels, thematic repetitions, methodological similarities, and presentational alikeness in AI art are hardly surprising since artmaking inevitably entails some degree of obvious or implied creative processing of artistic and other cultural artefacts. Various modes of this creative processing have been validated in different ways throughout 20th-century art, from Cubism and Dada, through Pop-Art, Fluxus, and Conceptual Art, to Postmodernism.[4] Widely accepted and most recognizable as part of the remix culture,[5] artefactual creativity[6] permeates all contemporary art disciplines and plays an important role in exploratory applications of computation for transforming existing data, ideas, relations, and phenomena.

However, artefactual creativity involves a deceptively smooth continuum of procedures ranging from interpretation, free copy, reprise, remake, allusion, citation, dedication, derivation and détournement, through mashup, remix, pastiche, reference, reminiscence, homage, and parody, to imitation, plagiarism, and forgery.[7] Expressive values of this procedural realm unfold in a grey zone of

---

**1** Cetinić / She 2022, Browne 2022.
**2** Grba 2022.
**3** Grba 2023.
**4** Haber n.d.
**5** Navas / Gallagher / burrough 2016.
**6** Artefactual creativity, also called (re)creativity Lessig 2006, is the application of combinatorial inventiveness to the specific qualities, meanings, contexts, or implications of existing artefacts to produce interesting new artefacts Grba 2015, Grba 2020, p. 74.
**7** Boon 2013, Grba 2015.

cultural inertias, dispersed knowledge, subtle influences, fuzzy ethical notions, and slippery moral categories, which fundamentally relativize the concept of (and to some degree the requirement for) authenticity or originality.[8] Furthermore, the expressive undercurrents, tendencies, and trends are closely interwoven with the fabrics of artists' professional lives and can be hard to identify and resist. Within such context, the grayest aspect of artefactual creativity concerns the imitation of another creation's poetic arc that synergizes a particular idea or topic with the most suitable, economic, or elegant elements such as formal rendering, production technique, or presentation format. Although the variational repetition of this synergetic relationship is unavoidable and necessary for apprenticeship in many areas of creativity, in artmaking it establishes the artwork's core identity and can be considered exploitative if the new artwork does not enhance the imitated poetic arc with meaningful values. The abstractness of these meaningful values sometimes makes it difficult to define the exploitation objectively.

For all these reasons, the assessment of expressive similarities navigates a fine and often blurry line of distinction meandering around fraudulent, flawed, and proper artistic strategies and always risks turning out as hasty, biased, uninformed, nitpicking, or preachy. Nevertheless, when an apparent similarity of relevant creative factors or a strong poetic parallel between a new artwork and a reasonably knowable referent remains unacknowledged by the new artwork's creator, comparative criticism is legitimate. It is invaluable for the maturation of AI art, whose originality-related mishaps often indicate deeper related issues and challenges caused by its techno-cultural entanglements and other factors.

AI art and its superset computational art have a long history of poetic similarities imposed by cognitive requirements, such as procedural literacy,[9] or induced by common methodologies. By their coding skills, AI artists range from creative directors who employ programmers or collaborate with computer scientists, through *bricoleurs* and tinkerers with various types and levels of technical expertise, to artist-engineers/scientists. They usually appropriate and sometimes modify the existing software packages or code libraries[10] and train them with publicly available datasets, which leads to aesthetic homogeneity. Attempting to break out of it, some artists pursue authenticity through obsessive experimentation with code,[11] thus revealing the lack of appreciation that originality is highly contextual, so its overidentification with technical or formal properties is largely misconceived.[12] Such superficial approaches relate to a more problematic type of expressive normalization that results from coupling AI systems with simplistic conceptual frameworks.

Formal repetitiveness, trivial contextualization, and shallow conceptualization are evident in Inceptionism, GANism, AI-derivative mainstream art, spectacular large-scale installations, and other subdomains.[13] Crucially, regardless of their causes, many AI artworks' similarities remain undisclosed by their authors. The prevalence of these "phantom" or "orphan" references merits a detailed study that would substantially exceed this paper's available volume, so we must be content here with a few multifaceted exemplars.

For the short film *Sunspring*,[14] Ross Goodwin trained one ML system on 162 science fiction (SF) movie scripts found online to generate the screenplay and screen directions, and another one on a folk songs database to generate the

---

**8** Although many poetic similarities are intentional (rip-offs), some result from (a combination of) enthusiasm, ignorance, carelessness, disregard, overconfidence, or negligence, which makes it difficult to categorize them with certainty.
**9** Procedural literacy is the ability to think, read, and write processes to engage procedural representation and aesthetics. It is closely associated with programming skills Reas / McWilliams / LUST 2010.
**10** For instance, the widely-discussed case of the Obvious' *Portrait of Edmond Belamy* (2018) laid the ground for questions about the nature of authenticity, the "localization" of expressive agency (in humans or AI systems), the creative autonomy, and the relativity of property liabilities in AI art McCormack / Giffors / Hutchings 2019, pp. 1, 3, Browne 2022, Audry 2021, pp. 164, 250 and Cetinić / She 2022, pp. 10–11.
**11** Miller 2019.
**12** Saltz 2019.
**13** Grba 2022, p. 19.
**14** Directed by Oscar Sharp, 2016.

song lyrics.[15] Oscar Sharp used this material to produce the film. Brimming with plot inconsistencies and awkward dialogues, *Sunspring* touches upon several issues of its underlying cultures. The artists' satirical application of ML to filmmaking reverses the logic of corporate movie search algorithms, playfully mimics Hollywood's screenwriting paradigm of regurgitating themes from earlier films, and anticipates the current application of ML for screenplay design. It simultaneously exemplifies the power and the perils of using statistics to trace the "cloud" of common ideational threads in a specific cultural domain. *Sunspring*'s incongruity when compared with conventional SF narratives also offers an analogy for the frivolity or nonsensicality of SF imaginaries in real life.

Two years later, Alexander Reben appropriated Hollywood strategies and regurgitated *Sunspring*'s concept and methodology to produce *Five Dollars Can Save the Planet* (2018)— "the world's first TED talk written by an AI." The text of this 3-minute TEDx talk was generated by training an ML model on "all TED talks."[16] Aiming to join the ongoing critique of TED's intellectual sharing model,[17] this work's humorous take on the trend of "robotization" of sales-pitch public lectures echoes Doug Zongker's more radical comic act *Chicken Chicken Chicken* (2007)—a 4-minute lecture in which the whole narration and different types of slides contain one word: Chicken.[18] Moreover, Reben's satirical logic and production methodology duplicate Goodwin and Sharp's while his choice of auto-recursive format (critiquing TED talks in a TED talk) mirrors Benjamin Bratton's 2013 TEDx talk *New Perspectives: What's Wrong with TED Talks?* [19] Although *Sunspring* is conceptually akin to SF parodies such as *Dark Star* [20] and its implication that the palatability of popular expressive forms partly relies on clichés nods toward Jennifer and Kevin McCoy's works with pop-cultural sampling,[21] its creators authentically activated one of the SF tropes—artificial intelligence—to make these points "mathematically." *Five Dollars…* uses an identical approach for a parodic statement about corporate public talks, but reveals or adds nothing new.[22]

Libby Heaney's two-channel video *Elvis* (2019) further illustrates the delicate dependency between an artwork's conceptual, topical, or methodological authenticity and critical cogency. Featuring a portrait of Heaney deepfaked as Elvis Presley and Presley's portrait deepfaked as Heaney, it directly copies (but does not acknowledge) the emblematic Gavin Turk's *POP* (since 1993).[23] In a series of selfie-pop-icon chimeras, *POP* addresses the same topics of individual identity and cultural mechanisms of celebrity mythmaking, involves the same pop icon, and applies the same formal method (face swapping) albeit in different media (sculpture, photographs, and prints) and a more complex chain of artistic/cultural allusions (acknowledged by Turk): For instance, a figure of Sid Vicious with Turk's face posing as Andy Warhol's *Elvis Presley* (1963). The sole diversion in Heaney's *Elvis* is the introduction of an AI technique (deepfaking) into the tactical repertoire of gender construction within digital technologies. Even in this aspect, *Elvis* is not too convincing when compared with the earlier uses of digital technologies to question the visual aspects of gender construction, such as image manipulations by Inez van Lamsweerde in the mid-1990s (V.A 1995).[24]

## 3. Conclusion

To whatever degree the persuasive weight of liminal differences in *Sunspring* vs. *Five Dollars…*, *Elvis* vs. *POP*, and numerous other cases[25] may be considered an open question or a matter of individual interpretation, poetic

---

**15** Goodwin 2016.
**16** Reben 2018.
**17** Morozov 2012, Harouni 2014.
**18** Bauman 2007.
**19** Bratton 2013.
**20** Directed by John Carpenter, 1974.
**21** See, for example, McCoys' *Every Shot, Every Episode* (2001) McCoy / McCoy 2023a and *Every Anvil* (2002) McCoy / McCoy 2023b.
**22** The conceptual and methodological cloning of *Sunspring* continued on a Gesamtkunstwerk level with the project *Legend of Wrong Mountain* (2018). Its central part is a generative video of a traditional Chinese *Kunqu* opera produced by a team of computer engineers, artists, and designers who trained an assortment of ML models on four different datasets about the forms of *Kunqu* opera to make the script (libretto), musical score, gesture choreography, and a woodcut book Huang 2019.
**23** Heaney 2019, Turk n.d.
**24** Brusatin / Clair 1995.
**25** For further consideration of questionable expressive similarities in AI art, whose listing similarities profoundly affect AI art. They signal a nonchalance toward both legacy and current creative landscapes, which is perhaps the most frustrating weakness of modern AI art and one of the key aspects for its constructive critique. We can hardly attribute it to the field's youth because AI art is more than 50 years old and shares all major poetic features with experimental arts whose history reaches back to at least the late 19th century.

Contemporary AI art's diverse affordances for addressing a vibrant ecosystem of important topics should make expressive inertias superfluous. Yet, "incidental reverberations" persist not only as accidental byproducts of the spontaneous convergence of ideas,[26] cognitive requirements, or technical constraints but also due to less justifiable psychological factors such as carelessness, indolence, egoism, arrogance, narcissism, and vanity mixed with professional deficiencies such as inexperience, contextual ignorance, or sketchy art-historical knowledge. They happen when artists disregard the essential requirement for multidirectional questioning and iterative critique of conceptual cogency, thematic relevance, and formal soundness in their creative processes. The artists' lack of transparency about their "phantom references" degrades AI art's images in the eyes of the general audience, which is chronically suspicious of the flimsiness, flippancy, and widely (mis)perceived lack of authenticity in contemporary art.

Some artworks, such as Disnovation.org's *Predictive Art Bot* (since 2017)[27] or Adam Basanta's *All We'd Ever Need Is One Another* (2018),[28] facetiously leverage the formal ambiguities of originality and authorship in modern ML to broach the vague terrain of ideational homogeneity in AI art and popular culture, expose the professional priorities of technical merit and monetary gain, and challenge the worldviews and interests that shape the intellectual property regulation in the digital age.[29] Their critique underlines that curious, earnest, and respectful exploration of general and field-specific art history is the basic—and empowering—artistic method for bringing up new works responsibly.[30]

But such tactical provocations are exceptional. The specific cases of AI art's unacknowledged poetic similarities and ethically charged slippages remain seldom exposed and openly discussed in the professional community probably because its members prefer to stay out of the reputational minefield that opens by clearly expressing potentially confrontational opinions. This self-protecting leniency goes in tandem with the equally present, but even less discussed, meritocratic inconsistency imposed by cultural hegemonies, power games, and systemic injustices of the contemporary art world. Indicating darker shades of human nature, it retains the accidents of birth, nationality, language, or geographical location as important vectors of career trajectories and recognition.

Along with other creative contingencies,[31] poetic similarities place AI art firmly within the modern context of existential crises, cultural contradictions, and societal tensions. They help us infer artists' knowledge and skills in light of their personal qualities (which inform poetics as much as any other expressive factor) and reaffirm that understanding artists equally as creators and human beings should be integral to the art appraisal. They provide an invaluable perspective for studying the strengths and deficiencies of AI art and for articulating the critical discussion of art and creativity in general.

---

would exceed this paper's available volume, see Grba 2023, p. 213.
**26** The somewhat analogous legal category of "independent creation" applies to cases when an existing creative expression gets reproduced without awareness of the original, so the new expression may not legally constitute a copy (Zeilinger 2021).
**27** Maigret / Roszkowska 2017.
**28** Basanta 2018.
**29** See the discussion of the *Predictive Art Bot* in Grba 2021, p. 247 and the discussion of *All We'd Ever Need Is One Another* in Zeilinger 2021, pp. 94–108.
**30** Grba 2023, pp. 217–218.
**31** See Grba 2023.